\documentclass[10pt]{article}
\usepackage{amsmath,amssymb}
\batchmode

\newtheorem{thm}{Theorem}

\newcommand{\pr}{\noindent{\bf Proof}. }
\newcommand{\re}{\noindent{\bf Remark}. }

\newcommand{\hs}{ \hspace{1cm}}

\newcommand{\al}{\alpha}
\newcommand{\De}{\Delta}
\newcommand{\de}{\delta}

\newcommand{\La}{\Lambda}

\newcommand{\Om}{\Omega}
\newcommand{\om}{\omega}

\newcommand{\cC}{{\cal C}}
\newcommand{\cD}{{\cal D}}

\newcommand{\cH}{{\cal H}}
\newcommand{\cS}{{\cal S}}

\newcommand{\cF}{{\cal F}}

\newcommand{\bbR}{{\mathbb{R}}}
\newcommand{\bbZ}{{\mathbb{Z}}}
\newcommand{\bbC}{{\mathbb{C}}}

\begin{document}

\title{The Dirac  sea}
\author{ 
J. Dimock  \footnote{email:  dimock@buffalo.edu}\\
Dept. of Mathematics \\
SUNY at Buffalo \\
Buffalo, NY 14260 }
\maketitle

\begin{abstract}
We  give  an alternate  definition   of the free Dirac field  featuring  an explicit construction of the  Dirac  sea.
The treatment employs  a  semi-infinite wedge  product of Hilbert spaces.  We also show that the construction is 
equivalent to the standard Fock space construction.
\end{abstract}

\section{Introduction}

Dirac  invented the Dirac equation  to  provide a  first  order relativistic differential  equation for the electron       which allowed a
quantum mechanical  interpretation.  He succeeded in this goal but there was  difficulty with the 
presence of solutions   with arbitrarily   negative energy.   These could not be excluded when the  particle
interacts with radiation and represented a serious instability.     Dirac's  resolution of the problem was to
to  assume  the particles were fermions,  invoke the Pauli exclusion principle,  and   hypothesize
that   the negative energy states  were  present   but  they were  all  filled.    The  resulting 
sea of particles  (the Dirac  sea)   would be stable and   homogeneous and   its  presence would not ordinarily be detected.
However  it  would be  possible  to  have some  holes in the sea which would behave as
if they had positive energy  and opposite charge.  These would be identified with anti-particles.
If   a  positive   energy   particle    fell into the sea and filled the hole  (with  an accompanying  the  emission of   photons), it would look as though     the particle  and anti-particle   annihilated.      The resulting picture  is known as  hole theory.  It gained credence with the discovery of   the positron,  the anti-particle of the electron.

The full interpretation of the Dirac  equation  came  with  the development of quantum field theory.
This is   a  multi-particle  theory  and  solutions of the Dirac equation are promoted to  quantum field operators. 
In  this framework it  is convenient   to  abandon the  hole theory   and  introduce the anti-particles as  separate 
entities.   Today   hole  theory  is mostly   regarded  as  inessential  and possibly misleading -   see the introduction  in  Weinberg   \cite{Wei95}.

Still the idea  retains a certain raw appeal  and it seems like a good idea to  keep our options open.
A  difficulty  with taking hole theory  seriously  is  that a satisfactory  mathematical  framework   
has   apparently  not been  developed in detail.   The purpose of this paper is to fill  this gap  by giving a 
construction  of the Dirac field operator based on hole theory.   The basic  idea is  that  if an $n$-fermion state is modeled by  an  $n$-fold  wedge product of Hilbert spaces,   then the Dirac sea
should  be descibed  as  an  infinite    wedge product of Hilbert spaces.

The treatment is not entirely  original.   We  take over  a similar construction which   has  been  used in 
the study of  infinite  dimension   Lie  algebras     \cite{Fei84},  \cite{Kac90}   and       has    
found applications  in string theory  \cite{FGZ86}.

\section{Semi-infinite wedge product}

We   start with a  complex  infinite dimensional  Hilbert space  $\cH$  which has   
a  fixed   decomposition  into two  (infinite dimensional)  subspaces
\begin{equation}
\cH   =  \cH^+  \oplus  \cH^-
\end{equation}
We   choose an  orthonormal basis      $\{e_i\}$   for  $\cH$   indexed by  $\bbZ - \{0\} $
which is compatible with  the splitting in the sense that 
$ e_1, e_2,  e_3, \dots  $  is a basis  for   $\cH^+$    and 
   $e_{-1},  e_{-2}, e_{-3},  \dots$ is a basis   for  $\cH^-$.   Let    $I$  be  a sequence
of non-zero integers  
\begin{equation}
I  =   ( i_1,  i_2, i_3, \dots  )      
\end{equation}
such  that      
\begin{equation}
  i_1 > i_2 > i_3 > \dots  
  \end{equation} 
and such  that for $k$ sufficiently large  $i_{k+1}   =  i_k  -1$.      The  set of all such sequences is a countable  set.
Associated with each  sequence  define  a formal symbol
\begin{equation}  \label{formal}
e_{I}  =   e_{i_1}  \wedge  e_{i_2}  \wedge  e_{i_3}   \dots  
\end{equation}
The semi-infinite  wedge  product  $\La_{\infty}$  is    the complex  vector space  of  all formal linear combinations
\begin{equation}
\sum_{I}   c_I   e_{I} \hs   c_I  \in \bbC
\end{equation}
with  $c_I  =0$  except for finitely many  $I$.      We  define an  inner product on this space  by taking the $e_I$ 
as an orthonormal basis.    Thus   $(e_I,  e_J)  =  \de_{IJ}$  and in general 
\begin{equation}
\Big(   \sum_I  c_I e_I,     \sum_J  c_J e_J
\Big)    =  \sum_I   |c_I|^2
\end{equation}
The completion  of  $\La_{\infty}$  in the associated norm  is  a Hilbert space   $\cH(\La_{\infty})$.

We    define  interior and  exterior  multiplication on $\La_{\infty}$  by  
\begin{equation}
\psi(e_j)  \Big(   e_{i_1}  \wedge  e_{i_2}  \wedge  \dots  \Big)
=  
\begin{cases}
0       &   \textrm{ if }  j \neq   i_s  \textrm{ for all  } s  \\
(-1)^{s+1} e_{i_1}  \wedge  e_{i_2}  \wedge  \dots  \wedge e_{i_{s-1}}  \wedge  e_{i_{s+1}}  \wedge  \dots   
   &   \textrm{ if }  j  =   i_s  \textrm{ for some   } s  \\
\end{cases}
\end{equation}
and   
\begin{equation}
\psi^*(e_j)  \Big(   e_{i_1}  \wedge  e_{i_2}  \wedge  \dots  \Big)
=  
\begin{cases}
0       &   \textrm{ if }    j  =   i_s  \textrm{ for some   } s   \\
(-1)^{s} e_{i_1}  \wedge  e_{i_2}  \wedge  \dots  \wedge e_{i_s} \wedge e_j \wedge  e_{i_{s+1}}  \wedge  \dots   
   &    \textrm{  if   }   i_s> j> i_{s+1}\\
\end{cases}
\end{equation}
These  are  adjoint to each other and  have the anti-commutators
\begin{equation}
\{\psi(e_i), \psi^*(e_j)  \}  =  \de_{ij}  \hs   \{\psi(e_i), \psi(e_j)  \}=0    \hs   \{\psi^*(e_i), \psi^*(e_j)  \}=0 
\end{equation} 
From the first it follows  that  for   $\Psi  \in   \La_{\infty}$
\begin{equation}
  \|  \psi^*(e_j) \Psi  \|^2    +  \|  \psi(e_j) \Psi  \|^2     =   \| \Psi  \|^2
\end{equation}
Hence   $\|  \psi(e_j) \Psi  \|  \leq   \| \Psi  \|$     
so   $ \psi(e_j) $   extends to a bounded operator on   $\cH(\La_{\infty})$  as does the adjoint.

There is   a distinguished state  $\Om_{\cD}$   defined by  
\begin{equation}   \label{sea}
\Om_{\cD}   =    e_{-1}  \wedge  e_{-2}  \wedge  e_{-3}  \wedge \dots    
\end{equation}
and  we  have   
\begin{equation}
\begin{split}
\psi(e_j)  \Om_{\cD}  =&  0      \hs     j>0  \\
\psi^*(e_j)  \Om_{\cD}  =&  0      \hs     j<0  \\
\end{split}
\end{equation}

To  complete the interpretation  of  $\cH(\La_{\infty})$  as   a semi-infinite   wedge product of  Hilbert spaces    we  need to    define
interior   and  exterior products for  all  $f \in  \cH$   and  to show that the construction is independent
of the choice of basis.   This is the content of the  next two theorems.

\begin{thm}
For   $f \in   \cH$  the sums   
\begin{equation}  \label{def}
\begin{split}
\psi(f)   =&  \sum_{i \in  \bbZ - \{0\} }   (f, e_i)   \psi(e_i)\\
\psi^*(f)   =&  \sum_{i \in  \bbZ - \{0\} }  (e_i,f)   \psi^*(e_i)\\
\end{split}
\end{equation}
converge in  operator  norm   to bounded operators  on    $\cH(\La_{\infty})$.  They  satisfy  
\begin{equation}     \label{anti}
\{\psi(f_1),  \psi^*(f_2)  \}    =   (f_1,f_2) 
\end{equation} 
with all other anti-commutators equal to zero.  Furthermore
 \begin{equation}     \label{bounds}
 \|  \psi(f)  \|  \leq   \| f\|      \hs     \|  \psi^*(f)  \|  \leq   \| f\|   
 \end{equation}   
\end{thm}
\bigskip

\pr    First  sum the sum  in    (\ref{def}) is finite.     
Then   $ \psi(f)$   and  $\psi^*(f)$   are adjoint to each other and satisfy  
\begin{equation}
\begin{split}
\{\psi(f_1),  \psi^*(f_2) \}  =&\sum_{ij}      (f_1,e_i)  (e_j,f_2)   \{\psi(e_i), \psi^*(e_j)  \} 
= \sum_{i}        (f_1,e_i)  (e_i,f_2)   =  (f_1,f_2)  \\
\end{split}
\end{equation} 
It   follows  that 
\begin{equation}
   \|  \psi^*(f) \Psi  \|^2   +\|  \psi(f) \Psi  \|^2     = \|f\|^2   \| \Psi  \|^2
\end{equation}
which implies the bounds   (\ref{bounds})  in this case.   

For  general  $f$   take the finite   approximation    
\begin{equation}
f_N = \sum_{|i|  \leq  N}  (f, e_i)  e_i   \hs   \psi(f_N)   =  \sum_{|i|  \leq  N}  (f, e_i)  \psi( e_i )    
\end{equation}
Then  $f_N  \to  f$ in  $\cH$  and so  
 \begin{equation}
\|  \psi(f_N) - \psi (f_M)  \|  =    \|  \psi(f_N-f_M)  \|  \leq   \|f_N  - f_M \|    \to   0
\end{equation}     
  as   $N,M  \to \infty$.   Hence   $\psi(f)  =  \lim_{N \to  \infty}  \psi(f_N)$
exists  and   similarly for the adjoint.        The  identity  (\ref{anti})     and  the bound   (\ref{bounds}) 
follow  by taking limits.
\bigskip

\re  Note that  $\psi(f)$ is anti-linear in $f$  (our convention is that  $(f,e_i)$ in anti-linear in $f$)
and   $\psi^*(f)$ is linear in $f$.  Also note that  
\begin{equation}  \label{die}
\begin{split}
\psi(h)  \Om_{\cD}  =&  0      \hs     h \in \cH^+  \\
\psi^*(g)  \Om_{\cD}  =&  0      \hs     g  \in \cH^-  \\
\end{split}
\end{equation}
Furthermore  liner combinations of   vectors  of the form   
\begin{equation}  \label{lc}
\prod_{i=1}^n   \psi^*(h_i)   \prod_{j=1}^m   \psi(g_j)  \Om_{\cD}  \hs    h_i  \in \cH^+,  g_j  \in \cH^-
\end{equation}
are dense   since they    include all vectors
of the form  (\ref{formal}).  \bigskip

\begin{thm}  \label{early}
The triple  $   \cH(\La_{\infty}),  \psi (f),  \Om_{\cD} $
constructed from a  basis  $\{ e_i\}$ compatible with the splitting  $\cH  = \cH^+  \oplus  \cH^-$    is independent of the basis  
in the sense that if     $  \cH(\La'_{\infty}),  \psi' (f),  \Om'_{\cD}$
is  a triple constructed from another basis   $\{ e'_i\}$  compatible with the splitting, 
then   there is a unitary operator  $U:    \cH(\La_{\infty}) \to   \cH(\La'_{\infty})$
such that  
\begin{equation}
\begin{split}
U \Om_{\cD}   =&   \Om'_{\cD}  \\
U \psi(f) U^{-1}  =&  \psi'(f)  \\
\end{split}
\end{equation}
\end{thm}
\bigskip

\pr     Consider   vectors of the form   (\ref{lc}).
  We have the inner  product   
\begin{equation}    \label{inner}
\begin{split}
&\Big(   \prod_{i=1}^{n_1}   \psi^*(h_{1,i})   \prod_{j=1}^{m_1}   \psi(g_{1,j})  \Om_{\cD}, \   
 \prod_{k=1}^{n_2}   \psi^*(h_{2,k})   \prod_{\ell=1}^{m_2}   \psi(g_{2,\ell})  \Om_{\cD}  \Big)  \\
=&  \left( \sum_{\pi}  \textrm{sgn}(\pi) \prod_{i=1}^{n_1} (h_{1,i},  h_{2,\pi(i)}) \right)
 \left(  \sum_{\pi'}  \textrm{sgn}(\pi') \prod_{j=1}^{m_1} (g_{1,j},  g_{2,\pi'(j)}) \right)
  \de_{n_1,n_2}    \de_{m_1, m_2}   \\ 
\end{split}
\end{equation}
where  $\pi$ is the permutations of  $(1, \dots,  n_1)$  and $\pi'$  is the permutations
of   $(1, \dots,  n'_1)$.
This follows  by  first   moving all  $\psi^*(h_{1,i})$    on the left  to the other  side of the inner product 
where they become  $\psi(h_{1,i})$.    Then  continue moving the  
 $\psi(h_{1,i})$  to the right   and   move  the  $\psi^*(h_{2,k})$  to the left using the anti-commutation
 relations   (\ref{anti}).   When they  hit   $ \prod_j  \psi(g_{1,j})  \Om_{\cD}$
or   $ \prod_{\ell}   \psi(g_{2,\ell})  \Om_{\cD} $  they give  zero.   If  $n_1 \neq  n_2$ there
are no surviving terms,  while if  $n_1 =n_2$  we  get the indicated sum over  $\pi$.  
Now  give a similar   argument with the surviving  
  $\left( \prod_j  \psi(g_{1,j})  \Om_{\cD},  \prod_{\ell}   \psi(g_{2,\ell})  \Om_{\cD} \right)$
  to get the sum over  $\pi'$.

We  define  $U$  on finite linear combinations of such  vectors    by  
\begin{equation}
U  \Big(  \sum_{\al}    \prod_{i=1}^{n_{\al}}   \psi^*(h_{ \al,i})   \prod_{j=1}^{m_{\al} }  \psi(g_{\al,j})  \Om_{\cD}  \Big)
= \sum_{\al}   \prod_{i=1}^{n_{\al}}   \psi'^*(h_{ \al,i})   \prod_{j=1}^{m_{\al} }  \psi'(g_{\al,j})  \Om'_{\cD}  
\end{equation}
This is inner product preserving by   (\ref{inner}),  hence it sends a zero sum to a zero sum,
 hence  the mapping is independent of the representation,  and so it  is well-defined.
Since it is norm preserving with dense domain and dense range it extends to a unitary.

\section{ Dirac equation}
We  review some   standard facts about the  Dirac equation.  (See   for example \cite{Dim11}).
The   Dirac equation  for a $\bbC^4$ valued  function  $\psi=  \psi (t, x)$ on  $\bbR \times  \bbR^3$
has  the form  
\begin{equation}
i \frac{d}{dt}  \psi   = H \psi  \equiv    (-i \nabla  \cdot \alpha  +  \beta  m  )  \psi
\end{equation}
where  $ \al ^1,  \al^2,  \al^3,  \beta$  are   self-adjoint  $ 4 \times  4$  matrices satisfying
\begin{equation}
\{\al^i,  \al^j\}  = 2 \de^{ij}  \hs   \{ \al^k, \beta\}  =  0  \hs   \beta^2  =  I
\end{equation}
The  Dirac Hamiltonian $H$ is self-adjoint   on a suitable domain in    the Hilbert space   
\begin{equation}
\cH =  L^2(\bbR^3, \bbC^4)
\end{equation}  
and  the solution to the equation is  $\psi(t,x)
=  (e^{-iHt} \psi)(x)$.       The spectrum of   $H$  is  $(-\infty, m]  \cup  [ m, \infty)$  and 
there is a corresponding  splitting of the  Hilbert  space   into positive and negative 
energy subspaces 
\begin{equation}
\cH  = \cH^+  \oplus  \cH^-
\end{equation}
With respect to this  splitting  the Hamiltonian has the form 
\begin{equation}
H  =  \om   \oplus  (- \om  )   \hs    \om  = \sqrt  {  - \De  +m  }
\end{equation}
It  is the positive energy subspace which  gives the states   of a single free particle  and 
the time evolution  for  such states is    $\psi(t,x) =  (e^{-i\om t} \psi)(x)$. 
All the above   statements are best established  by going to momentum space  with the  Fourier transform.

We  note  also that     the projection  onto   $\cH^{\pm}$  is  given by  
\begin{equation}
P^{\pm}   =  \frac{ \om \pm   H}{   2 \om  }
\end{equation}
 In  addition there is an anti-linear  charge conjugation operator  $\cC$   on  $\cH$   such that 
 $\cC^2 =I$  and   $( \cC \psi,  \cC \chi)  =  ( \chi, \psi)$.  It maps  $\cH^{\pm}$ to $\cH^{\mp}$  and 
 satisfies  $\cC P^{\pm} =  P^{\mp}  \cC  $.

\section{Quantization on the Dirac Sea}

The Dirac  field  operator   should  be  a solution  $\psi(t,x)$ of the Dirac equation taking values in
the bounded operators on  
some  complex Hilbert space  such that  the intial  field  $\psi(x)  =  \psi(0,x)$
satisfies    the anti-comutation relations  $\{\psi_{\al}(x) ,  \psi_{\beta}^*(y) \}  =  \de(x-y) \de_{\al \beta}$.  
These  requirements    should be interpreted in  sense  of distributions at least 
in the spatial variable.  
 Thus  for  a function  $f$  in the Schwartz space   $\cS(\bbR^3,  \bbC^4)$
 we  ask  for  field operators   $\psi(t,f) $   (formally   $\sum_{\al =1}^4\int \psi_{\al}(t,x) \overline {f_{\al}(x)} dx$) 
 which are anti-linear  in  $f$  and  satisfy  
 \begin{equation}    \label{field}
i \frac{d}{dt}  \psi(t,f)   =  \psi(t, Hf) 
\end{equation}
with  intial  conditions    $\psi(f)  =  \psi(0,f)$   satisfying
\begin{equation}  \label{anti2}
\{\psi(f_1) ,  \psi^*(f_2) \}  = (f_1,f_2)
\end{equation}
In   addition we would like time evolution to be unitarily implemented with
positive energy.   That is there should be a positive self adjoint operator  $H'$ 
such that   
\begin{equation}
 \psi(t,f) =e^{iH't} \psi(f)  e^{-iH't}  
\end{equation}

Having  constructed   the semi-infinite tensor product  the  solution is now easy. 

\begin{thm}     \label{main}
Let    $\cH(\La_{\infty} ), \psi(f),  \Om_{\cD}$  be  the semi-infinite wedge product  defined    for  the splitting
$\cH = \cH^+  \oplus  \cH^-$   into positive and negative energy.  
Then  the   field operator  
\begin{equation}
\psi(t,f)   \equiv      \psi(  e^{iHt} f)    
\end{equation}
has the anti-commutator   (\ref{anti2})   and  satifies the field equation  (\ref{field}).
Time evolution is unitarily implementable with positive energy.
\end{thm}

The verification  of the first  two points is immediate;  the derivative can even be taken
in norm thanks to (\ref{bounds}).  We  postpone the unitary implementability.
\bigskip

Note that if we split     $f=P^+ f  +  P^-f$ the  field operator   can also be written
\begin{equation}    \label{sting}
\psi(t,f)  =    \psi(  e^{i\om t}P^+ f)  +       \psi(  e^{-i\om t}P^- f)  
\end{equation}

Next we  explore the particle content of this structure.       The state $\Om_{\cD}$ defined in (\ref{sea})   is 
the Dirac sea filled with negative energy particles .     The general  state is obtained by
applying field operators  $\psi(f)$ to  $\Om_{\cD}$  as in  (\ref{lc}),  and these  differ only locally
from  $\Om_{\cD}$.  The  operators    $ \psi^*(e_i), \psi(e_i)$  for $i>0$  create or annihilate      positive energy particles  
and  the same is  true  for     $ \psi^*(h),  \psi(h)$   if   $h \in \cH^+$.
Therefore we  define particle creation and annihilation operators by    
\begin{equation}  
a^*(h) =  \psi^*(h) \hs    a(h) =  \psi(h)  \hs    h \in \cH^+
\end{equation}
 The  operators    $ \psi^*(e_i), \psi(e_i)$  for $i<0$  create or annihilate   negative energy particles  in the sea
and  the same is  true  for     $ \psi^*(g),  \psi(g)$   if   $g \in \cH^-$.
According to the  hole theory picture we    want  to  regard the annihilation of  a  negative energy particle   
as the  creation of a positive energy   anti-particle of opposite charge,  and the  creation of a negative energy particle 
as   the annhilation of an positive energy  anti-particle of opposite charge.   Therefore  we    define anti-particle   creation and annihilation  operators    by   
\begin{equation}  
b^*(h) =  \psi( \cC h) \hs    b(h) =  \psi^*(\cC h)  \hs    h \in \cH^+,  \cC h  \in \cH^-
\end{equation}
Then  $a^*(h), b^*(h)$ are linear in $h$ while   $a(h), b(h)$  are anti-linear in $h$.  We
  have the anti-commutation relations  
\begin{equation}
\{a(h_1),  a^*(h_2) \}  =  (h_1,h_2)    \hs  \{b(h_1),  b^*(h_2) \}  =  (h_1,h_2)  
\end{equation}
with all other anti-commutators equal to zero.
Furthermore   (\ref{die}) becomes   
\begin{equation}
a(h) \Om_{\cD}=0  \hs   b(h) \Om_{\cD}=0
\end{equation}
and   applying   $a^*(h), b^*(h)$ to    $\Om_{\cD}$  generates a dense set.

Now  we  rewrite the field operator.  
Note that   $ \psi(  e^{-i\om t}P^- f)   =  b^*( \cC e^{-i\om t}P^- f)$ which can also be written 
$ b^*(  e^{i\om t} \cC P^- f) $    or    $ b^*(  e^{i\om t}  P^+\cC f) $.
Thus the field operator  (\ref{sting}) can  be written
 \begin{equation}    \label{sting2}
\psi(t,f)  =   a(  e^{i\om t}P^+ f)  +      b^*(  e^{i\om t}  \cC P^- f)  
\end{equation}
The field annihilates particles and creates  anti-particles.
The adjoint 
 \begin{equation}    \label{sting3}
\psi^*(t,f)  =   a^*(  e^{i\om t}P^+ f)  +      b(  e^{i\om t}  \cC P^- f)  
\end{equation}
 creates particles and annihilates anti-particles.

 \section{Quantization on Fock Space - a comparison}
 
 We   show  that  our  quantization  is equivalent to the usual  quantization of Fock space  in
 which particles and anti-particles are introduced as separate particles.
 (See   for example \cite{Dim11}).    Starting with  the positive energy subspace  
 $\cH^+$  of    $\cH  =  L^2(\bbR^3, \bbC^4)$
 let   $\cH^+_n$   be the $n$-fold anti-symmetric product  (wedge product)  and let
   $ \cF(\cH^+) = \oplus_{n=0}^{\infty} \cH^+_n$   be the associated fermion Fock space.    Further let  $\al(h), \al^*(h)$ be the standard creation  and annihilation operators for  $h  \in \cH^+$.  If   $\Om_0$
 is  the no particle  state then  $\al(h) \Om_0 =0$.

 The full Hilbert space  is  a tensor product    
 \begin{equation}
 \cF  =  \cF(\cH^+)  \otimes   \cF(\cH^+)
\end{equation} 
 The first  factor is the particle  Fock space and the second factor is this anti-particle Fock space.
 Creation and annhilation operators for particles and anti-particles   are given by 
 \begin{equation}
 a(h)   =  \al(h)  \otimes I  \hs     b(h)   = (-1)^N  \otimes   \al(h) 
\end{equation}
and their adjoints.
 We again have the anti-commutation relations  
\begin{equation}
\{a(h_1),  a^*(h_2) \}  =  (h_1,h_2)    \hs  \{b(h_1),  b^*(h_2) \}  =  (h_1,h_2)  
\end{equation}
and thanks to  the factor  $(-1)^N$  all other anti-commutators are zero.
If       $\Om_0 =  \Om_0  \otimes  \Om_0$  is  the no particle  state  in   $\cF$
then    $a(h) \Om_0=  b(h) \Om_0 = 0$  and applying  $a^*(h),  b^*(h)$  to  $\Om_0$
generates a dense set.

Now   the  time  zero    Dirac field operator can be defined as a distribution by  
\begin{equation}
\psi(f)  =a( P^+ f)   +  b^*(\cC  P^-  f)
\end{equation}
This   has the anti-commutator    (\ref{anti2})  and the  time evolution    
 $\psi(t,f)  =  \psi(e^{iHt} f)$   satisfies the field equation (\ref{field}).    It  can also be written   
 \begin{equation}
 \psi(t, f)  =a(e^{i\om t  }   P^+ f)   +  b^*(e^{i \om t }\cC  P^-  f)
 \end{equation}
which has the same form  as  (\ref{sting2}).   

 In  the Fock space representation it is well-known that    time  evolution is unitarily 
implementable with positive energy -  the Hamiltonian  $H'$  is just 
the multi-particle version   of $\om$.    In  the final  theorem  we  establish that the Fock space  construction is unitarily
 equivalent   to the Dirac sea construction.    Then     time evolution is unitarily 
 implementable  with positive energy  for the Dirac sea  as well,  and the proof of 
 theorem  \ref{main}    is complete.

 \begin{thm}  
 There is  a unitary operator   $U:  \cH(\La_{\infty})  \to  \cF$  such that 
 \begin{equation}
 \begin{split}
 U  \Om_{\cD}   =&  \Om_0  \\
 U   a(h)  U^{-1}   =& a(h)   \hs    U   b(h)  U^{-1}  =b(h)  \\
 \end{split}
 \end{equation}
 and  hence   
 \begin{equation}
  U   \psi(t,h)  U^{-1}   = \psi(t,h)  
 \end{equation}
\end{thm}
\bigskip

\re  On the left side of these equations  $a(h), b(h), \psi(t,f)$  refer to the Dirac sea operators
and on the right side they refer to the Fock space operators.
\bigskip

 \pr  The idea of the proof is the same as  the proof of theorem  \ref{early}.
    Consider   vectors of the form  
\begin{equation}  \label{vector2}
\prod_{i=1}^n   a^*(h_i)   \prod_{j=1}^m  b^*(h'_j)  \Om_{\cD}  \hs    h_i, h_j'  \in \cH^+
\end{equation}
or the same  with  $\Om_0$  instead of  $\Om_{\cD}$ in the Fock representation.
 Using the anti-commutations relations   and  the fact that  $a(h), b(h)$  annihilate 
 the   $\Om_{\cD}$ or  $\Om_0$  we  have  the   have the inner  product   
\begin{equation}    \label{inner2}
\begin{split}
&\Big(   \prod_{i=1}^{n_1}   a^*(h_{1,i})   \prod_{j=1}^{m_1}  b^*(h'_{1,j})  \Om_{\cD}, \   
 \prod_{k=1}^{n_2}   a^*(h_{2,k})   \prod_{\ell=1}^{m_2}   b^*(h'_{2,\ell})  \Om_{\cD}  \Big)  \\
=&    \left(  \sum_{\pi}\textrm{sgn}(\pi) \prod_{i=1}^{n_1} (h_{1,i},  h_{2,\pi(i)}) \right)  \left( 
\sum_{\pi'} \textrm{sgn}(\pi')  \prod_{j=1}^{m_1}  (h'_{1,j},  h'_{2,\pi'(j)}) \right) \  \de_{n_1,n_2}
\de_{m_1, m_2}   \\ 
\end{split}
\end{equation}
 and  exactly the same in the Fock representation.

 Now  define
  $U$  on  linear combinations of such vectors    by  
\begin{equation}
U  \Big(  \sum_{\al}   \prod_{i=1}^{n_{\al}}   a^*(h_{ \al,i})   \prod_{j=1}^{m_{\al}}   b^*(h'_{ \al,j})  \Om_{\cD}  \Big)
= \sum_{\al}   \prod_{i=1}^{n_{\al}}   a^*(h_{ \al,i})   \prod_{j=1}^{m_{\al}}   b^*(h'_{\al,j})  \Om_0  
\end{equation}
This is inner product preserving by   (\ref{inner2}),  hence it sends a zero sum to a zero sum,
 hence  the mapping is independent of the representation,  and so it  is well-defined.
Since it is norm preserving with dense domain and dense range it extends to a unitary.

\end{document}